\documentclass[journal, twoside]{IEEEtran}
\usepackage{indentfirst}
\usepackage{graphicx}
\usepackage{amsmath}
\usepackage{amssymb}
\usepackage{amsfonts}
\usepackage{mathrsfs}
\usepackage{leftidx}
\usepackage{color}
\usepackage{amsmath}
\usepackage{arydshln}
\usepackage{amsthm}
\usepackage{ragged2e}
\usepackage{cite}
\usepackage{enumerate}
\usepackage{longtable}
\usepackage{float}
\usepackage{stfloats}
\usepackage{hyperref}
\usepackage{algpseudocode}
\usepackage{algorithm}
\usepackage[caption=false,font=footnotesize]{subfig}
\usepackage{multicol} 
\usepackage{makecell}
\usepackage{url}
\usepackage{enumitem}
\usepackage{multirow} 
\usepackage{booktabs}
\theoremstyle{plain}

\usepackage{caption}

\newcolumntype{P}[1]{>{\raggedright\arraybackslash\footnotesize}m{#1}}
\newcolumntype{A}[1]{>{\centering\arraybackslash\footnotesize}m{#1}}

\usepackage[table,usenames,dvipsnames]{xcolor}
\definecolor{aa}{RGB}{175,238,238}
\definecolor{bb}{RGB}{255,255,255}

\usepackage{bm}
\usepackage{makecell}

\begin{document}

\title{Important Bits Prefix M-ary Quadrature Amplitude Modulation for Semantic Communications}

\author{Haonan Lu, Rui Meng,~\IEEEmembership{Member,~IEEE,} Xiaodong Xu,~\IEEEmembership{Senior Member,~IEEE,} 
Yiming Liu,~\IEEEmembership{Member,~IEEE,} 

Ping Zhang,~\IEEEmembership{Fellow,~IEEE,} and Dusit Niyato,~\IEEEmembership{Fellow,~IEEE}

\thanks{

This work was supported in part by the National Key Research and Development Program of China under Grant 2020YFB1806905; in part by the National Natural Science Foundation of China under Grant 62501066 and under Grant U24B20131; and in part by the Major Science and Technology Program of Hebei Province under Grant 202621502010001.

Haonan Lu, Rui Meng (\textit{Corresponding author}, E-mail: buptmengrui@bupt.edu.cn), Xiaodong Xu, Yiming Liu, and Ping Zhang are with State Key Laboratory of Networking and Switching Technology, Beijing University of Posts and Telecommunications, Beijing 100876, China.
Rui Meng and Xiaodong Xu are also with the Satellite Internet Testing Center, Xiong'an Aerospace Information Research Institute, Xiong'an 070001, China.
Dusit Niyato is with College of Computing and Data Science, Nanyang Technological University, Singapore.

The code is available at https://github.com/mouren21/IBP
}}

\maketitle

\begin{abstract}
M-ary Quadrature Amplitude Modulation (MQAM) is a commonly used channel modulation technology in wireless communication systems. To achieve dedicated channel modulation for semantic communication (SemCom), we propose an Important Bits Prefix MQAM (IBP-MQAM) scheme and derive its approximate expression of important symbol error rate (ISER) and unimportant symbol error rate (USER). By extracting and quantifying text semantics using Dynamic Proximal Policy Optimization (DPPO) and Latent Dirichlet Allocation (LDA), we verify that IBP-MQAM achieves improved performance over MQAM in SemCom scenarios and further analyze the effects of key system parameters.

\end{abstract}


\section{Introduction}

Through the deep integration of artificial intelligence and wireless communication technologies, semantic communication (SemCom) has emerged as a highly promising paradigm. Unlike conventional systems, which prioritize the transmission of raw bit sequences, SemCom focuses on conveying the desired semantic content and aligning communication with task objectives, offering substantial efficiency gains \cite{9955312,cheng2026apeg,fan2026generative}.

Most existing SemCom schemes emphasize joint source and channel coding (JSCC) strategies, focusing on improving semantic fidelity \cite{10584091, 10251411,10960697,10521803}. For instance, \cite{10960697} introduces a universal JSCC framework to mitigate the influence of channel variations, and \cite{10521803} presents a bit allocation strategy based on dynamic proximal policy optimization (DPPO) to enhance bit utilization efficiency. To further  investigate channel effects, \cite{10854563} analyzes symbol error rate (SER) and semantic distortion under conventional modulation. However, conventional modulation techniques do not consider the special characteristics of different bit importance in SemCom. Therefore, some recent studies attempt to tailor modulation techniques specifically for SemCom \cite{10978695,10979950}. For example, \cite{10978695} proposes representing semantic features using Gaussian distributions, and \cite{10979950} introduces a semantic-oriented modulation scheme that maps semantic information (SemInf) directly into discrete vectors, achieving favorable resource usage. \cite{wang2025semantic} proposes SOM and \cite{teng2025sdmcm} proposes sDMCM, both of which focus more on moving from a lower bit error rate (BER) to a lower MSE the debugging process. The features are mapped to a symbol, so they are not inflexible for semantic compression. Current research on unequal error protection (UEP) for SemCom primarily relies on error-correcting code\cite{10426244,10571004} or channel resource allocation based on channel state information\cite{6490257}. Traditional hierarchical modulation (HM) schemes are mostly designed to approach the Shannon limit of the overall BER, rather than to optimize for UEP requirements. 
Against this background, we propose a channel modulation scheme for SemCom. The key contributions are summarized as follows:

\begin{itemize}
    \item We propose the Important Bits Prefix M-ary Quadrature Amplitude Modulation (IBP-MQAM) scheme to address the limitation of conventional MQAM in leveraging the semantic relevance of transmitted information. The proposed IBP-MQAM scheme maintains compatibility with MQAM while significantly improving task performance in SemCom.
    
    \item We provide two ways of distinguishing important bits and unimportant bits. We further derive analytical expressions for SER, important SER (ISER), and unimportant SER (USER) under the proposed IBP-MQAM and conventional MQAM schemes. 
    
    \item 
    We use Monte Carlo simulation to verify the derived theoretical formula. Simulation results on open-source datasets \cite{LANG1995331,sanjid_hasan_swandip_snigdha_2025,coates2011analysis} verify that, the derived approximate expressions are basically consistent with the simulation results; IBP-MQAM has better task accuracy than MQAM under SemCom; and IBP-MQAM achieves higher cosine similarity (CosSim) than MQAM in SemCom. 
\end{itemize}

\section{Proposed IBP-MQAM Scheme}
\subsection{System Model}

\begin{figure}[t]
\centering
\includegraphics[width=3in]{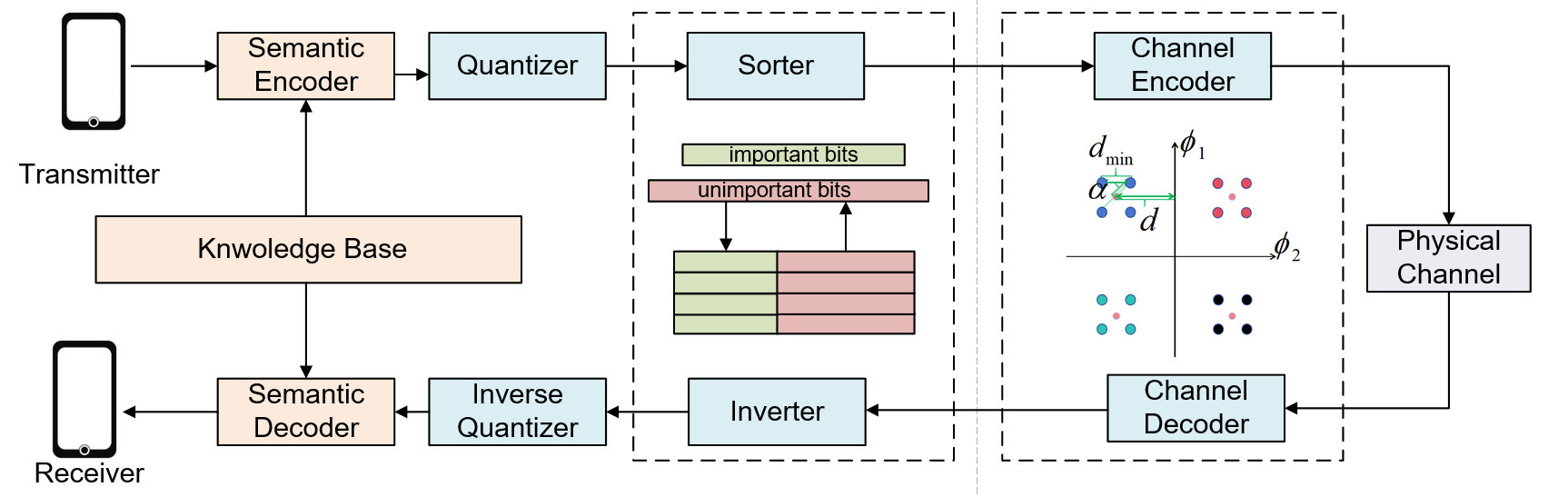}
\caption{SemCom framework based on the proposed IBP-MQAM constellation structure. Taking QPSK as an example, $\alpha$ adjusts the spacing between suffix symbols $d_{min}$ and their respective QPSK center $d$.}

\label{model}
\end{figure}

As illustrated in Figure \ref{model}, we consider a SemCom system, where information-bearing features are sparsely encoded and transmitted over a noisy wireless channel using the proposed IBP-MQAM scheme. At the transmitter, the semantic encoder, assisted by the knowledge base, generates SemInf, which is quantized into important and unimportant bits. These are reordered, modulated by the channel encoder, and then transmitted. At the receiver, the channel decoder demodulates the symbols, the bits are recovered and restructured, and the SemInf is reconstructed via inverse quantization and the knowledge base \cite{10854563}. Each modulated symbol comprises a fixed-length bit sequence of $n$ bits, where the first $p$ bits ($p \ll n$) are designated as the prefix bits, encoding the most critical SemInf. The remaining $s = n - p$ bits form the suffix bits, which contains less significant details.

Considering the fact that the proportion of important bits generated by the model is too large and some systems do not commonly use channel modulation methods with too large modulation indices, we focus on the case $p = 2$ in subsequent discussions\footnote{The selection of $p$ depends on the demand for compressing the number of symbols transmitted. In theory, the ratio of $p$ to $n$ is the high-importance SemInf and all SemInf, but a larger $p$ can transmit fewer symbols when transmitting the same number of bits, improving the utilization of bandwidth. When $p$ is an odd such as $3$, in order to avoid complex operations during demodulation and maintain the overall structural relationship, MQAM is also chosen.}, and map the prefix bits are mapped to the standard QPSK constellation $\mathcal{C}_p$ because of its superior performance according to \cite{s22166174}. 

Each QPSK point serves as the center of a sub-constellation defined by the suffix bits. As illustrated in Figure \ref{model}, IBP-MQAM introduces a scaling factor $\alpha \in [0,1]$ that controls the spread of suffix points around their corresponding QPSK centers. When $\alpha = 0$, all suffix-modulated symbols collapse onto the corresponding prefix constellation, effectively discarding suffix variability; When $\alpha = 1$, the modulation scheme becomes equivalent to a conventional $2^n$-QAM. This design enables flexible tradeoffs between important and unimportant bit error rates.
Additionally, gray coding is applied to the suffix bits along both in-phase and quadrature axes. All symbols sharing the same prefix reside within the same sub-constellation, centered around the associated QPSK point. The spacing between the constellation center and the coordinate axes is denoted by $d$, which serves as a reference for the symbol layout.

The transmitted passband signal can be expressed as
\begin{equation}
s(t) = X\phi_1(t) + Y\phi_2(t),
\end{equation}
where $X$ and $Y$ are the in-phase and quadrature components of the symbol, respectively.
$\phi_1(t) = \sqrt{\frac{2}{T}} \cos(2\pi f_c t)$ and $ \phi_2(t) = \sqrt{\frac{2}{T}} \sin(2\pi f_c t)$ are the orthonormal basis functions, where \(T\) denotes the symbol duration and \(f_c\) is the carrier frequency.
The signal passes through an additive white Gaussian noise (AWGN) channel, and the received baseband signal is
\begin{equation}
y(t) = s(t) + z(t),
\end{equation}
where $ z(t) \sim \mathcal{CN}(0, \sigma^2)$ denotes complex Gaussian noise with variance $\sigma^2$.
The demodulated components are obtained via correlation with the orthonormal basis functions as
\begin{equation}
X^{\prime} = \int y(t) \phi_1(t)\,dt = \frac{X + z_1}{2},
\end{equation}
\begin{equation}
Y^{\prime} = \int y(t) \phi_2(t)\,dt = \frac{Y + z_2}{2},
\end{equation}
where $z_1$ and $z_2$ are independent Gaussian noise projections along the two axes.

\subsection{Comparisons between MQAM and IBP-MQAM}

In this subsection, we analyze the protection properties of important bits under IBP-MQAM, focusing on the case where $\alpha = 1$, such that the proposed IBP-MQAM scheme becomes functionally equivalent to MQAM in terms of symbol set but retains a distinct bit-to-symbol mapping.

To analyze the bit protection capabilities, we define safe constellation points as those whose neighboring symbols differ only in unimportant bits. Conversely, $N$-level points denote constellation points having $N$ neighbors that differ in important bits. For a given modulation order $M$, the spatial arrangement of identical prefix bits in IBP-MQAM provides better spacing properties for important bits than in MQAM.
Furthermore, the number of safe and $N$-level points is summarized in Table~\ref{bit tybe} for typical values of $M$. The results indicate that the proposed IBP-MQAM yields more safe points and fewer 2-level points compared to MQAM, which means that IBP-MQAM has better protection of SemInf than MQAM.

\begin{table}[ht]
\centering
\caption{Comparison of Safe and $N$-Level Constellation Points}
\label{bit tybe}
\begin{tabular}{llll}
\toprule
\textbf{Bit Type} & \textbf{16QAM} & \textbf{MQAM($M>16$)} & \textbf{IBP-MQAM} \\
\midrule
Safe points & $0$& $\frac{M}{2} - 2\sqrt{M}$ & $(\sqrt{M} - 2)^2$ \\
1-level points & $8$ & $6\sqrt{M}$ & $4\sqrt{M} - 8$ \\
2-level points & $8$ & $\frac{M}{2} - 4\sqrt{M}$ & $4$ \\
\bottomrule
\end{tabular}
\end{table}

\subsection{Symbol Error Performance of IBP-MQAM and MQAM}

In this subsection, we adopt a geometric projection-based approach mentioned in \cite{9217986} that analyzes one coordinate axis at a time, to evaluate the symbol error performance of IBP-MQAM, where ISER is caused by errors in important bits (prefix bits), and USER results from errors in unimportant bits (suffix bits). 
Given that the basis functions are orthogonal, the error probabilities calculated along each axis can be directly combined multiplicatively.

Let $M' = 2^s$ represent the number of suffix constellation points along one axis. These constellation points are symmetrically distributed with respect to the origin, with $\sqrt{M'}/2$ points located on each side. Let $d_{\min}$ denote the minimum spacing between adjacent points on the same side. The relationship between $d$ and $d_{\min}$ is expressed as $d_{\min} = 4 d \alpha/\sqrt{M}$. The coordinates $x_i/ y_i$ that $X$/$Y$ can take are expressed as $d + \left(2i - 1 - \frac{\sqrt{M}}{2}\right) d_{\min}$, where $i \in \left\{1, 2, \dots, \frac{\sqrt{M}}{2}\right\}$. Using the sum of squares of odd integers: $\sum_{i=1}^{n}(2i-1)^2 = \frac{n(4n^2-1)}{3}$, the average energy of the positive half of the $X$-component is
\begin{equation}
\begin{aligned}
E_{Xh} &= \frac{2}{\sqrt{M}} \sum_{i=1}^{\frac{\sqrt{M}}{4}} \left[ \left(d + \frac{2i - 1}{2} d_{\min}\right)^2 + \left(d - \frac{2i - 1}{2} d_{\min}\right)^2 \right] \\
&= \frac{2}{\sqrt{M}} \sum_{i=1}^{\frac{\sqrt{M}}{4}} \left[ 2d^2 + \frac{2(2i - 1)^2 \cdot 4 d^2 \alpha^2}{M} \right] \\
&= \left(1 - \frac{4 \alpha^2}{3M} + \frac{\alpha^2}{3}\right) d^2.
\end{aligned}
\end{equation}
Since the distribution of $X$ is symmetric, the overall average energy $E_X$ equals $E_{Xh}$. The same holds for $Y$, and hence the total average symbol energy of IBP-MQAM is
\begin{equation}
E_s = \frac{E_X + E_Y}{2} = E_{Xh} = \left(1 - \frac{4 \alpha^2}{3M} + \frac{\alpha^2}{3}\right) d^2.
\end{equation}
We denote the distance between two constellation points as $D$, and assume the noise affecting the transmission is zero-mean with variance $\sigma^2$. Then, the probability that one constellation point is incorrectly decoded is
\begin{equation}
P = \int_D^\infty \frac{1}{\sqrt{2 \pi \sigma^2}} \exp\left(-\frac{v^2}{2 \sigma^2}\right) dv = Q\left(\frac{D}{\sigma}\right).
\end{equation}
In the projection model, ISER is dominated by confusion between points on opposite sides of the origin but with the same suffix bits. Therefore, the high-probability ISER per coordinate axis is approximated as
\begin{equation}
\begin{aligned}
P_{h-\text{I}} = \frac{2}{\sqrt{M}} \sum_{k=1}^{\frac{\sqrt{M}}{2}} Q\left( \frac{d - \frac{\sqrt{M} - (4k - 2)}{4} d_{\text{min}}}{\sigma} \right) \\= \frac{2}{\sqrt{M}} \sum_{k=1}^{\frac{\sqrt{M}}{2}} Q\left( \sqrt{\frac{\left( 1 - \alpha + (2k - 1)\frac{2\alpha}{\sqrt{M}} \right)^2 E_s}{\left( 2 - \frac{8\alpha^2}{3M} + \frac{2\alpha^2}{3} \right) \sigma^2}} \right)
\end{aligned}
\end{equation}
The decisions on the two coordinate axes do not affect each other, and the original symbol can only be restored when the decisions on both coordinate axes are correct. Thus, we have
\begin{equation}
P_{\text{I}} = 1 - (1 - P_{\text{h-I}})^2 = 2P_{\text{h-I}} - P_{\text{h-I}}^2.
\end{equation}
USER is governed by misclassifications within the same prefix group. By a similar analysis as ISER, it's given by (\ref{10}). We have
\begin{figure*}[!b] 
    \centering
    \scriptsize 
    \setlength{\abovedisplayskip}{0pt} 
    \setlength{\belowdisplayskip}{0pt} 
    \setlength{\jot}{-2pt} 
    \begin{equation} 
        \begin{split}
        P_{h-\text{U}} &=\Bigl( 2 - \frac{4}{\sqrt{M}} \Bigr) Q\Bigl( \frac{d_{\text{min}}}{2\sigma} \Bigr)  + \frac{2}{\sqrt{M}} \sum_{i=1}^{\frac{\sqrt{M}}{2}-1} \biggl[ Q\Bigl( \frac{2d - \frac{\sqrt{M} - 8i + 2}{4} d_{\text{min}}}{\sigma} \Bigr) - Q\Bigl( \frac{2d - \frac{\sqrt{M} - 8i - 2}{4} d_{\text{min}}}{\sigma} \Bigr) \biggr] \\
        &=\Bigl( 2 - \frac{4}{\sqrt{M}} \Bigr) Q\Bigl( \sqrt{\frac{4\alpha^2 E_s}{M \Bigl( 2 - \frac{8\alpha^2}{3M} + \frac{2\alpha^2}{3} \Bigr) \sigma^2}} \Bigr) + \frac{2}{\sqrt{M}} \sum_{i=1}^{\frac{\sqrt{M}}{2}-1} \biggl[ Q\Bigl( \sqrt{\frac{\Bigl( 2 + \frac{8i - 4\alpha}{\sqrt{M}} \Bigr)^2 E_s}{\Bigl( 2 - \frac{8\alpha^2}{3M} + \frac{2\alpha^2}{3} \Bigr) \sigma^2}} \Bigr) - Q\Bigl( \sqrt{\frac{\Bigl( 2 + \frac{8i}{\sqrt{M}} \Bigr)^2 E_s}{\Bigl( 2 - \frac{8\alpha^2}{3M} + \frac{2\alpha^2}{3} \Bigr) \sigma^2}} \Bigr) \biggr]
        \end{split}
        \label{10}
    \end{equation}
    \vspace{-1em} 
\end{figure*}
\begin{equation}
P_{\text{U-IBP}} = 2 P_{\text{h-U}} - P_{\text{h-U}}^2.
\end{equation}
Because USER and ISER are approximately independent at per coordinate axis, symbol error is important bits error add unimportant bits error while important bits right at per coordinate axis.  The overall SER of IBP-MQAM is given by
\begin{equation}
P_{\text{e-IBP}} \approx 1 - (1 - P_1)^2 = 2P_1 - P_1^2,
\end{equation}
where
$P_1 = P_{\text{h-U}}(1 - P_{\text{h-I}}) + P_{\text{h-I}},$
and the BER of important bit $P_{\text{i}}$ and unimportant bit $P_{\text{u}}$can be obtained simply by dividing by the number of bits contained in each symbol and then multiplying by the constant $K$. We assume that a symbol is assigned with $A$ important bits and B unimportant bits, and get the result of its normalized MSE and MAE as
\begin{equation}
\begin{split}
E(\text{MSE}) &= E(e_{\text{Quan}}^2) + E(e_{\text{Chan}}^2) \\
&= \frac{\Delta^2}{12} + \Delta^2 \left( \sum_{k=0}^{B-1} 4^k P_u + \sum_{k=B}^{A+B-1} 4^k P_i \right) \\
&= \frac{0.25 - P_u}{3 \cdot 4^{A+B-1}} + \frac{P_u - P_i}{3 \cdot 4^{A-1}} + \frac{4P_i}{3}
\end{split}
\end{equation}
\begin{equation}
\begin{split}
E(\text{MAE}) &\approx E(|e_{\text{Quan}}|) + E(|e_{\text{Chan}}|) \\
&\approx \int_{-\frac{\Delta}{2}}^{\frac{\Delta}{2}} \frac{|x|}{\Delta} dx + \sum_{k=0}^{B-1} (\Delta \cdot 2^k) P_u + \sum_{k=B}^{A+B-1} (\Delta \cdot 2^k) P_i \\
&= \frac{0.5}{2^{A+B}} + P_u (2^{1-A} - 2^{1-A-B}) + P_i (2 - 2^{1-A})
\end{split}
\label{MAE}
\end{equation}
\section{Simulation Parameters}

\subsection{Parameters of SemCom}

To validate the effectiveness of the proposed IBP-MQAM scheme, we employ the following two SemCom approaches under AWGN channel conditions.

\subsubsection{Latent Dirichlet Allocation (LDA)-based SemCom}

It employs LDA as the semantic codec with IEEE 754 standard quantization, following the framework established in \cite{10854563}.

\begin{itemize}
	\item \textbf{Semantic Encoder}: LDA serves as a generative probabilistic model to discover latent topics from text corpora. Given a collection of $P$ documents with vocabulary size $Q$, LDA assumes $K$ latent topics and derives the posterior distributions $P(z_k|d_p)$ (topic given document) and $P(w_q|z_k)$ (word given topic) through inference techniques like Gibbs sampling. The semantic features comprise contained topics $Z^*_p$ and topic distribution $\rho_p$, where $\rho_p = \sum_{k=1}^K [P(z_k|d_p) - 1/K]^2$ represents the dispersion degree of topic-related content.
	
	\item \textbf{Quantize and Sorter}: The semantic features are quantized using IEEE 754 standard quantization, with important bits corresponding to the exponent components that determine the scale of the topic distribution values. TABLE \ref{tab:semantic_bow} shows that transmitting only partial bits will reduce semantic accuracy.
	
\end{itemize}
\begin{table}[ht]
\centering
\caption{Semantic Word Bags Comparison with Low Bits}
\label{tab:semantic_bow}
\footnotesize 
\begin{tabular}{p{2.8cm} p{5.5cm}}
\toprule
\textbf{Item} & \textbf{Content} \\
\midrule
LDA-Generated Bag (Lossless Benchmark) & new, car, price, buy, don, just, think, science \\
\midrule
Bag after Low-bit Quantization & new, car, price, buy, sale, don, just, think, people, like, science, data \\
\bottomrule
\end{tabular}
\end{table}
\subsubsection{DPPO-based SemCom}
This approach utilizes ResNet-18 as the semantic codec and employs a DPPO-based agent for dynamic bit allocation.

\begin{itemize}
    \item \textbf{Semantic Encoder:} The ResNet-18 architecture extracts hierarchical semantics from input images. The semantic importance vector $\boldsymbol{\omega}$ is computed by combining the Semantics Task Relevance (STR) and Inter-Semantics Relevance (ISR) as $\omega_{k}=g_{k}\times v_{k}$ \cite{10521803}.

    \item \textbf{Agent \& Reward Design:} The bit allocation is formulated as an MDP. To address the sparsity of task feedback, we propose a \textit{Difference Reward} mechanism. Let $U_t = \mathcal{L}_t - \beta \mathcal{D}_t$ be the utility at step $t$, where $\mathcal{L}_t$ denotes the instantaneous classification accuracy and $\mathcal{D}_t$ is the weighted MSE of features. This design encourages the agent to maximize the marginal gain of every allocated bit. The state space includes semantic features, importance weights $\boldsymbol{\omega}$, remaining budget, and the theoretical error probabilities ($P_i, P_u$) of the IBP channel.

    \item \textbf{Adaptive $\alpha$ Optimization:} Unlike static modulation, we implement an adaptive update for the IBP scaling factor $\alpha$. In each training iteration, we compute the optimal $\alpha^*$ that minimizes the expected semantic distortion based on the current feature importance and theoretical SER derivation in (\ref{MAE}). We then apply an MAE update: $\alpha \leftarrow \tau \alpha^* + (1-\tau) \alpha$, ensuring stable convergence to the optimal geometry for varying SNRs.

    \item \textbf{Quantizer}: Employs non-subtractive uniform dithered quantization with dynamic range $\gamma$ and quantization spacing $\Delta_i = 2\gamma/M_i$, where $M_i = 2^{b_i}$.
\end{itemize}

\subsection{Monte Carlo Simulation Results}
\begin{figure}[t]
    \centering
    \includegraphics[width=3in]{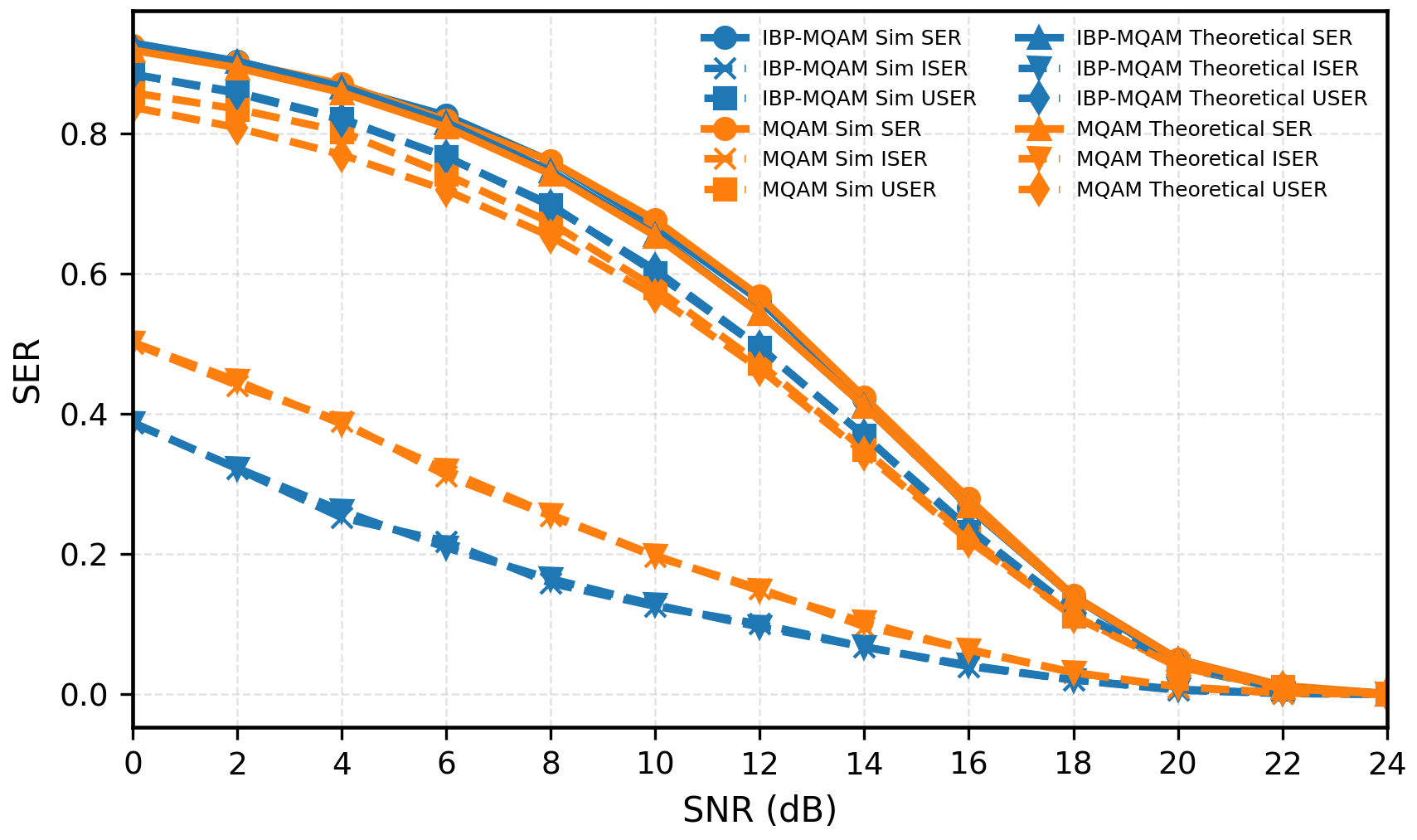}
    \caption{Simulation and theoretical results of SER, ISER, and USER for IBP-MQAM and MQAM.}
    \label{ser}
\end{figure}

As shown in Figure \ref{ser}, the simulation results closely match the theoretical derivations for SER, ISER, and USER when the SNR is relatively high. When the SNR is relatively low, errors in constellation points may span across multiple constellation points, causing some deviation between the theoretical results and simulations. Nevertheless, the general trend and relative relationship remain accurate. As observed, when $\alpha=1$, IBP-MQAM sacrifices some of the USER to achieve improved ISER performance while maintaining the same total SER as MQAM. This characteristic aligns with the original design intent of IBP-MQAM. Even at $\alpha=1$, with optimal bit allocation, IBP-MQAM can outperform MQAM.

\subsection{Dataset-based Simulation Results}

\subsubsection{Parameters of Open-source Dataset}
The LDA-based SemCom scheme uses 20 Newsgroup\cite{Adewoye2024ComprehensiveRO} and Bangladesh News Articles\cite{sanjid_hasan_swandip_snigdha_2025} datasets, with importance assigned to subject indices and topic components ($K=20$ topics, $\epsilon=0.01$ threshold, top-3 topics) to test the CosSim \cite{8947433} which is employed for joint evaluation of topic distribution and semantic feature recovery.

The DPPO-based SemCom scheme utilizes STL-10\cite{coates2011analysis} operating under a 1000-bit budget with 33\% important ratio to test the task accuracy which is the probability of successfully classifying correctly.

\begin{figure}[htbp]
    \centering
    \includegraphics[width=3in]{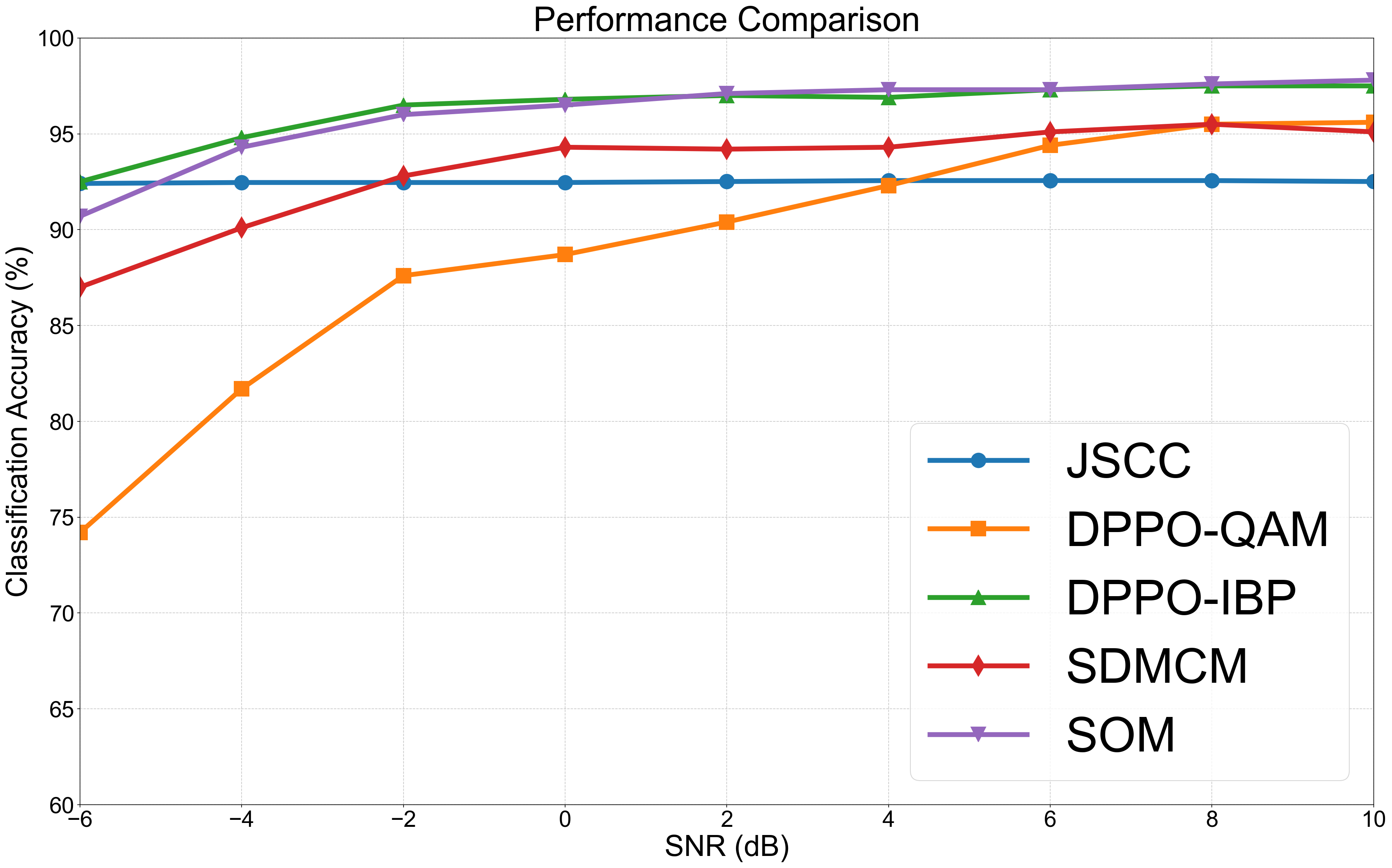}
    \caption{Task accuracy comparison between IBP-MQAM, MQAM, SOM\cite{wang2025semantic}, sDMCM\cite{xin2024semantic} and JSCC for STL-10 datasets. All schemes transmit 166 symbols, each symbol corresponding to 64 constellation points if there is a corresponding constellation point.}
    \label{task acc}
\end{figure}

\subsubsection{Task Accuracy vs. SNR}
As shown in Figure \ref{task acc}, all models are trained with 5 dB SNR, and the number of samples used for evaluation is 1e3. The JSCC in the figure has stabilized, and its 95\% confidence interval is 90.0\%-93.4\%. IBP-MQAM performs better than QAM across all evaluated SNR conditions. Within the typical SNR range of 0–5 dB, which represents common SemCom scenarios, IBP-MQAM consistently outperforms the SDMCM scheme. DPPO-IBP is better than SOM at 0 dB and below, performs similarly to SOM in 2–6 dB SNR, and slightly underperforms SOM at higher SNRs. SOM achieves stable performance as its quantization and noise errors do not superimpose, with no extra noise accumulation from cross-symbol bit allocation.  This observation underscores the effectiveness of the IBP strategy in enhancing the robustness and efficiency of SemCom systems under realistic channel conditions, positioning IBP-MQAM as a competitive candidate for practical deployment in semantic-aware communication tasks.

\begin{figure}[htbp]
    \centering
    \subfloat[$M=16$]{\includegraphics[width=0.24\textwidth]{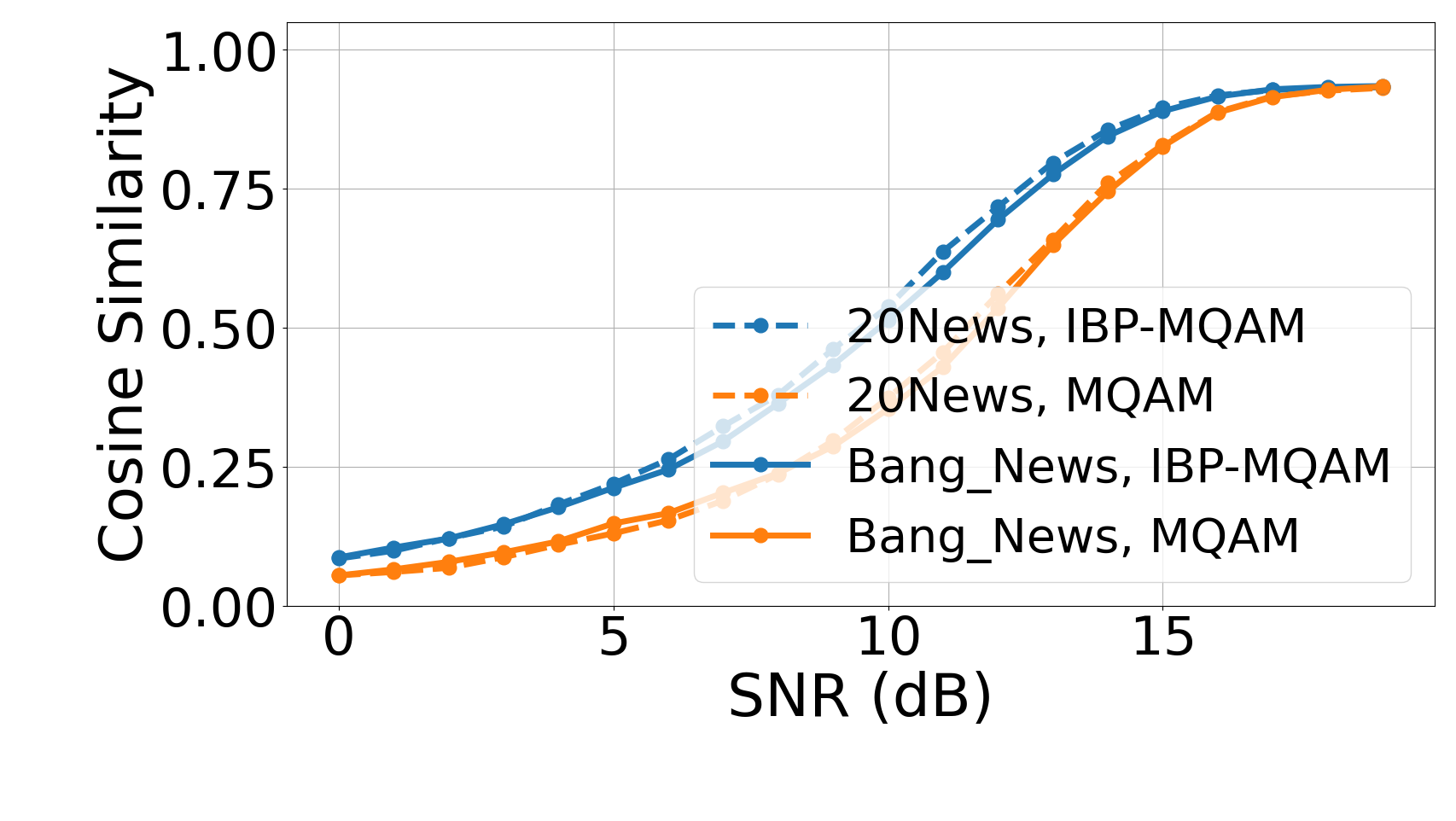}\label{cos16}} 
    \subfloat[$M=64$]{\includegraphics[width=0.24\textwidth]{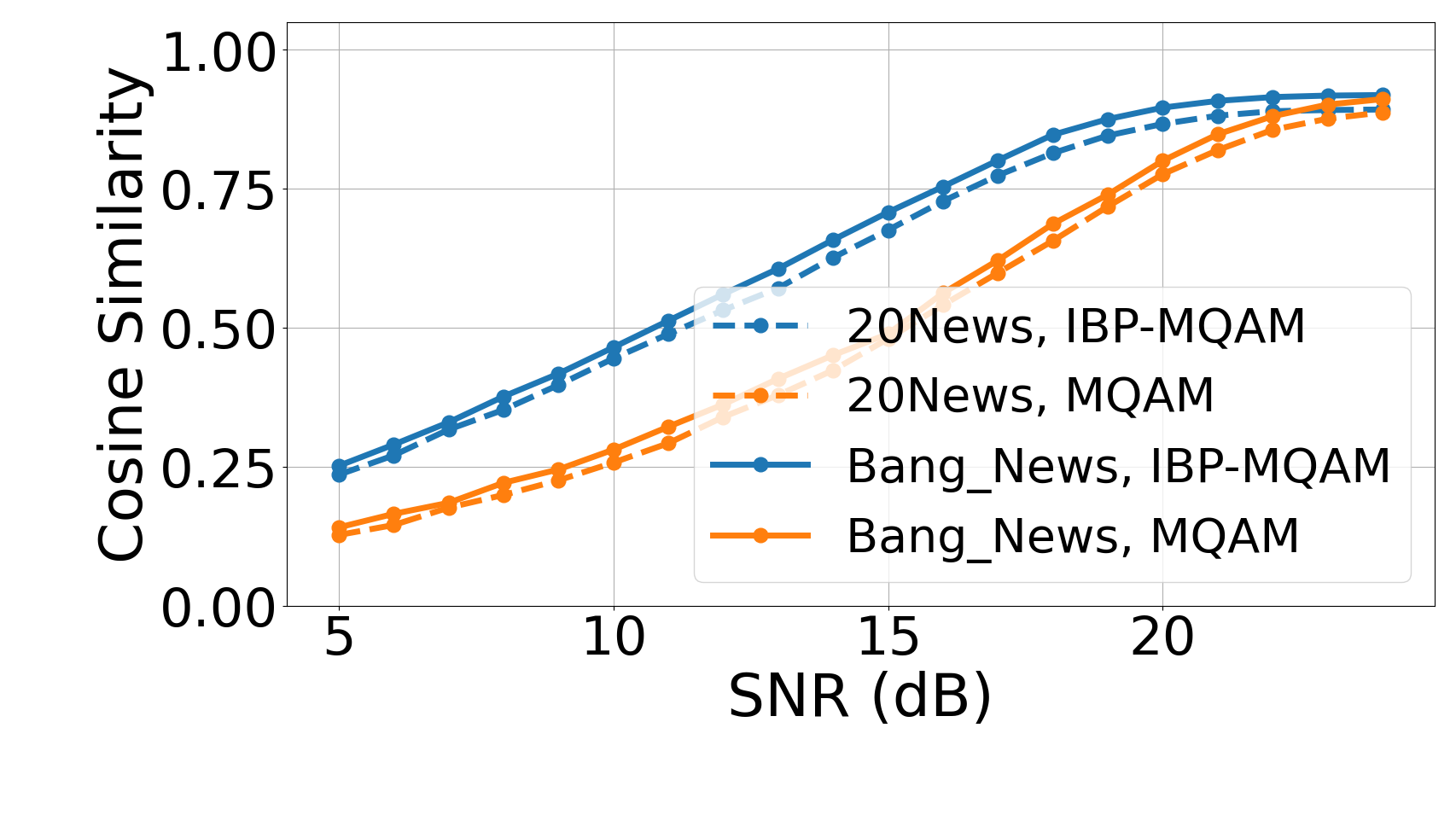}\label{cos64}} \\
    \caption{CosSim comparison between IBP-MQAM and conventional MQAM under varying modulation orders $M$.}
    \label{fig:cos}
\end{figure}

\subsubsection{CosSim vs. SNR}
Figure \ref{fig:cos} demonstrates that IBP-MQAM consistently outperforms conventional MQAM in terms of semantic transmission accuracy, across different modulation orders. This improvement aligns with our design rationale: by emphasizing the protection of important bits through constellation structuring and leveraging the compensation parameter $\alpha$, IBP-MQAM enhances robustness for semantically critical components while maintaining acceptable performance for non-critical bits. Meanwhile, we find that the performance improvement of the 20 Newsgroup dataset is more obvious than that of common-tests, and the curve IBP-MQAM is also smoother than that of MQAM, which shows that IBP-MQAM is more stable than MQAM, and it is easy to a better performance with a good model and a suitable dataset.

\subsubsection{Impact of $\alpha$}
\begin{figure}[t]
    \centering
    \includegraphics[width=1.7in]{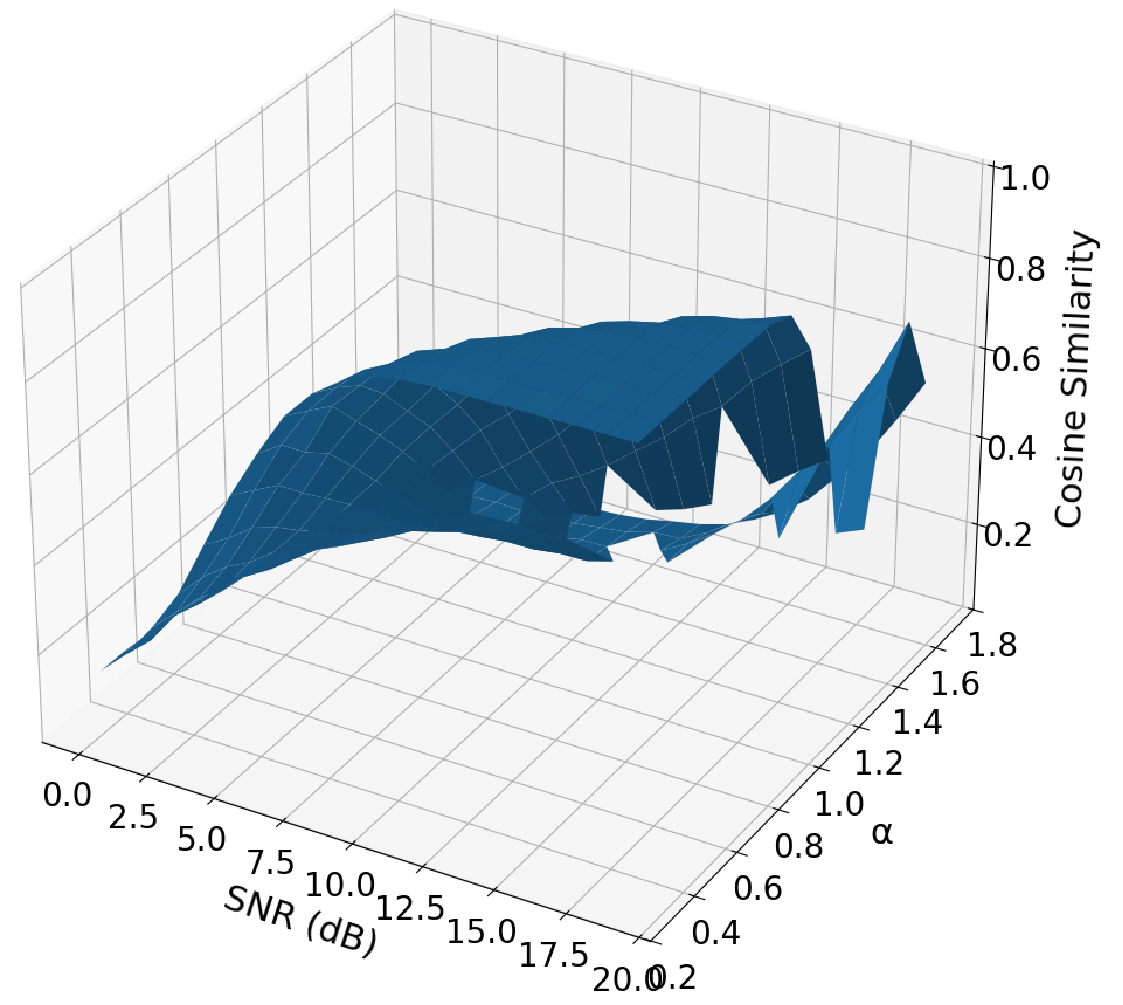}
    \caption{CosSim versus SNR under varying values of $\alpha$ for IBP-MQAM schemes.}
    \label{alphasnr}
\end{figure}

As illustrated in Figure \ref{alphasnr}, the influence of $\alpha$ on task accuracy is complex. As $\alpha$ decreases, the average symbol energy reduces for the same $d$, leading to increased separation between important bits and higher security for these bits. Conversely, as $\alpha$ increases, the spacing between symbols of the same important bit expands, improving the security of unimportant bits. At high SNR, CosSim fluctuates significantly when $\alpha$ is either too small or too large. A small $\alpha$ increases challenges for unimportant bits, while a large $\alpha$ causes overlap in constellation diagrams, resulting in interference. The impact of $\alpha$ is also dataset- and model-dependent, allowing for adjustment based on specific task requirements.
\section{Conclusion}

We have proposed the IBP-MQAM scheme to differentiate the protection levels for important and unimportant bits in SemCom systems. Through the tunable parameter $\alpha$, IBP-MQAM has enabled flexible modulation strategies to optimize transmission reliability for semantically
critical information. Simulation results have validated its effectiveness in enhancing task accuracy and stability compared to conventional MQAM. In the future, we will employ a hierarchical design to IBP-MQAM for multi-level semantic importance division, and further explore the fusion of IBP-MQAM with other semantic modulation schemes (e.g. sDMCM) in unimportant bit sub-constellations, to realize more flexible and efficient semantic transmission
for complex SemCom scenarios with richer knowledge base and more complex models.

\[\]
\bibliographystyle{IEEEtran}
\bibliography{ref}
\end{document}